\documentclass[conference]{IEEEtran}
\usepackage{amsmath}  

\usepackage{amsfonts} 

\ifCLASSINFOpdf
\else
\fi
\usepackage{graphicx}
\usepackage{algorithm}
\usepackage{algpseudocode}
\usepackage{amsmath} 

\begin{document}

%
\title{Bringing computation to the data: A MOEA-driven approach for optimising data processing in the context of the SKA and SRCNet}

\author{\IEEEauthorblockN{Manuel Parra-Royón, Álvaro Rodríguez-Gallardo, Susana Sánchez-Expósito, Laura Darriba-Pol, Jesús Sánchez-Castañeda,\\ MÁngeles Mendoza, Julián Garrido, Javier Moldón and Lourdes Verdes-Montenegro}
\IEEEauthorblockA{Instituto de Astrofísica de Andalucía, IAA-CSIC\\
Glorieta de la Astronomía, s/n\\
18018, Granada - Spain\\
Email: \{mparra, jmoldon, arodriguez\}@iaa.es}
}


%


\maketitle


\begin{abstract}
The Square Kilometre Array (SKA) is a next-generation radio astronomy-driven big data facility that will revolutionise our understanding of the Universe and the laws of fundamental physics, and needs innovative solutions for efficient data processing. The SKA Regional Centres Network (SRCNet) is a collaborative ecosystem tasked with the demanding role of processing and analyzing SKA data products. With SKA, the near-exascale computing will be a challenge, chief among them being the issue of data movement. As computational capabilities, the sheer volume of generated data becomes staggering. The traditional approach of moving tons of data to centralized computing resources becomes impractical due to the limitations of existing networks and storage infrastructures. The data transfer bottleneck becomes a critical impediment, hindering the overall efficiency. To overcome this challenge, a paradigm shift is imperative. Strategies such as in-situ processing and distributed computing models where computation is moved to the data emerge as promising solutions.

In the realm of SKA and specifically within SRCNet data processing needs, the conjunction of Function-as-a-Service (FaaS) with a decision-making entity driven by Evolutionary Algorithms (EAs) becomes pivotal. FaaS abstracts away infrastructure management concerns, enabling the deployment of modular functions in close proximity to data sources. This development aligns with the principle of bringing computation to the data, mitigating the challenges associated with extensive data transfers. The decision-making entity, guided by EAs, facilitates a systematic exploration of near-optimal execution plans, that will provide with detailed information on how and where a function should be executed within the overall computing and data infrastructure. With the focus on two objectives such as execution time and energy consumption, and constraints like data transfers or data locations, Multi-Objective Evolutionary Algorithms (MOEAs) could be a good option to optimise the movement of the computation to the data. In this context, MOEAs could provide a baseline guide for efficient and cost-effective data processing for the computation model within the SRCNet.
Our proposal unifies the technical aspects of FaaS deployment, together with the mathematical modelling and code implementation of a customised first approach MOEA model for the optimisation of function execution plans within the SRCNet architecture and its integration with FaaS.

\end{abstract}


%
\IEEEpeerreviewmaketitle

\section{Introduction}
The SKA telescopes will provide a significant leap in sensitivity, resolution and survey speed. Comprising two main arrays, SKA-Mid and SKA-Low, strategically located in South Africa and Western Australia respectively, the SKA will be able to span a broad frequency range from 50 MHz to 14 GHz. 
The SKA telescopes science goals are broad and ambitious ranging from the Cosmic Dawn to the Origin of Life, considering also exploratory studies. Due to the sheer volume of data it will have to transport, process, store and distribute to its end users around the globe, the SKA project is considered by many the ultimate Big Data challenge. 

The SKA Observatory (SKAO) will be supported by a global network of SKA Regional Centres (SRCNet), distributed around the world. The SRCNet \cite{garrido2023status} will play a pivotal role in advancing astronomical research by tailoring its data processing capabilities to the needs of the astronomers. The SRCNet will be strategically designed to efficiently archive, distribute and process the huge science data products generated by the SKAO telescopes. Approaches such as the SKA Data Challenges\cite{hartley2023ska,bonaldi2021square} provide insight into the volume and type of computing expected in the SRCNet.

\begin{figure}[ht]
  \centering
  \includegraphics[width=\columnwidth]{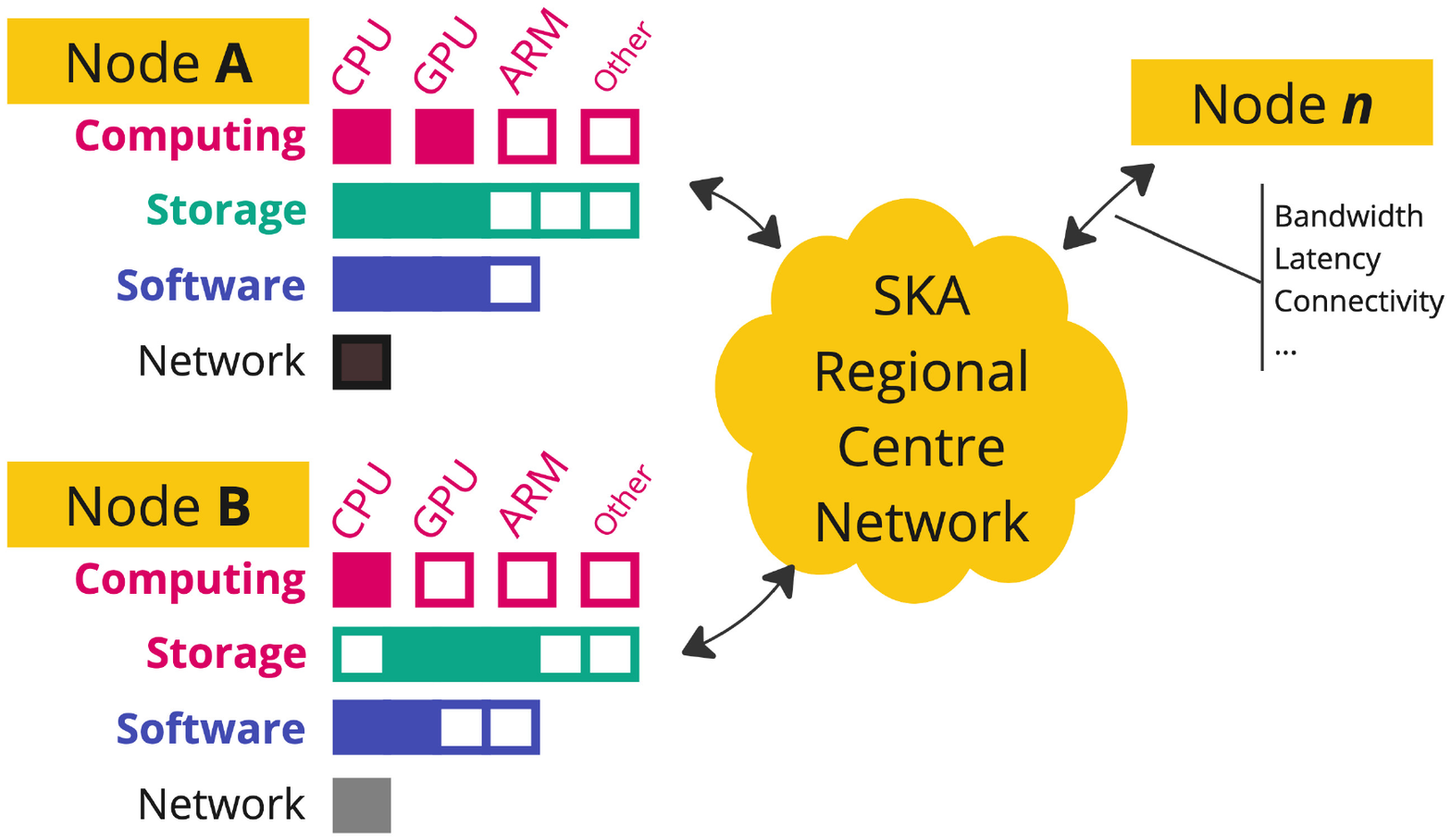}
  \caption{Diagram of a simplified SRCNet structure, showing different inter-connected nodes providing different computing elements, storage capacities, software/services availability and networking features. Each block represents a subset of elements, e.g., for Computing, CPU could be the number of available cores, Storage would be the available data collections, Software would refer to the availability of software for processing, and Network indicates connectivity with other nodes.}
  \label{fig:srcnet}
\end{figure}

The computing model within the SRCNet is yet to be defined. Several approaches have been proposed, ranging from cloud-based environments, distributed systems or data-mesh proposals. 
Scientific users will submit a wide variety of computing processes and tasks to the SRCNet, from complex pipelines running on top of workload managers (such as slurm) to isolated operations and functions focused on the use of large volumes of data. The data from the telescopes will be distributed globally across the SRCNet nodes, so it is necessary to consider where the data and computation are located so that the function to run is as close to the data as possible. This model defines a data-driven architecture where computation moves to where the data is \cite{machado2022data}. Diagram \ref{fig:srcnet}, shows a very simplified architecture of the SRCNet, where each of the nodes contributes to the system computing elements such as CPUs, GPUs, and others. Also, each node stores different collections of data, which may or may not be available at various locations or nodes, and deploys a set of software functions aligned with the computing capabilities. Networks also play an important role in determining part of the system constraints along with storage.

This architecture requires knowledge on the SRCNet such as the location of the data, permissions for accessing data and using computing resources, requirements from the computing tasks, software and hardware availability among other. Other parameters in the SRCNet, such as bandwidths between nodes or latency, should also be considered. In addition, it works according to the specifications of each suggested function execution. 

Function-as-a-Service (FaaS) presents a Cloud Computing (CC) \cite{rajan2020review} transformative paradigm that effectively mitigates substantial overheads associated with traditional computing models. By abstracting away infrastructure management tasks, FaaS\cite{jonas2019cloud}  allows developers to deploy individual functions independently, eliminating the need for continuous resource allocation. This streamlined approach significantly reduces time overhead, allowing fast and efficient execution of code directly on the data. Developers write specific functions, which are then deployed in a serverless environment on a CC provider. This approach eliminates the need for infrastructure management, allowing for a pay-as-you-go model and improving scalability as well as resource allocation. Although FaaS does not solve the whole problem, it provides an initial scenario that can accommodate the execution of tasks/functions over distributed data. In our case we can reduce the SRCNet problem to a model where users call individual functions that require computational resources such as CPU or GPU and that are applied on distributed data within SRCNet. These functions can be containerised, deployed  and served from a FaaS service. Platforms such as Kubernetes provide support for this, enabling features such as allocation of compute resources to specific functions, thus providing excellent granularity to accommodate highly scalable workloads. 




In this work, we propose a model design based on MOEAs to address the challenge of bringing computation to a distributed network of nodes storing collections of data in the most efficient and optimal way possible, considering aspects such as the capabilities of the nodes in terms of computing units (CPU, GPUs, etc.), storage or connectivity among others. The model focuses on two objectives: a) resource consumption/usage, b) energy efficiency of operations. The underlying idea is to derive an execution plan to determine how to execute tasks/functions on the data required by scientific users of the SRCNet. This  proposal takes into account the system's state to dynamically feed the MOEA and provide plans that adapt to the global execution demand context of the SRCNet.

The paper is structured as follows: Section 2 provides a summary of related works, subsequently, Section 3 develops the technical aspects concerning the deployed FaaS layer that will serve as orchestration facility for execution plans. In Section 4, we define the mathematical model and the implementation of the MOEA algorithm. Finally, in Section 5, we discuss the conclusions of the proposed model.

\section{Related works}

Many scientific works have explored the challenges and opportunities associated with computation and efficiency in distributed systems \cite{orgerie2014survey}. The complexities of optimising computational tasks across distributed nodes, emphasising the need for intelligent resource allocation are addressed in \cite{hussain2013survey}. Efficiency considerations extend beyond mere computational speed, encompassing factors like load balancing, fault tolerance, and adaptability to varying workloads. The work of \cite{yi2020task} contributes insights into achieving efficiency in distributed systems by optimizing task distribution and resource utilisation, laying the foundation for our approach to computational efficiency within the context of the SRCNet architecture.

Efficient management and processing of data in distributed systems have been also explored in the context of the Data Mesh paradigm \cite{habenicht2022syncmesh}. Data Mesh advocates for a decentralized approach to data architecture, aligning with the idea of bringing computation to the data. The decentralisation of data domains and the establishment of data products as independently deployable units resonate with the goals of optimizing data movement and enhancing overall system efficiency. This paradigm complements our strategy for orchestrating the execution of operations on distributed data within the SRCNet.

The advent FaaS has revolutionized CC paradigms. Works such as \cite{rajan2020review} and \cite{chadha2021architecture} explore the principles and applications of FaaS, highlighting its role in abstracting infrastructure complexities. FaaS enables the seamless deployment and execution of modular functions, presenting a paradigm shift in how developers interact with CC resources. Our work is in line with these works, incorporating FaaS as an  execution layer for orchestrating the utilisation of CC resources within the SRCNet.

MOEAs have gained prominence in addressing optimisation challenges, particularly in the realm of computation and efficiency. The work of \cite{ahmadi2022multi} demonstrates the effectiveness of MOEAs in finding optimal solutions for resource allocation and energy efficiency in distributed systems. Our work builds upon these foundations, employing an approach of MOEAs to dynamically optimize the execution plans for operations on distributed data in the SRCNet, balancing objectives such as resource consumption and  energy efficiency.

The integration of FaaS and MOEAs presents a novel approach to optimizing CC resources. In \cite{ma2021multi}, the synergy between serverless computing and evolutionary algorithms for resource-efficient applications is investigated. FaaS offers agility and scalability, while MOEAs provide a systematic way to balance conflicting objectives. Our proposed approach leverages this synergy, employing FaaS as an execution orchestration layer and MOEAs to dynamically adapt execution plans. 

Next section outlines the architecture and operational aspects of the FaaS layer, which serves as a dynamic orchestration mechanism for the execution plans.

\section{FaaS deployment}

\begin{figure}[h!]
  \centering
  \includegraphics[width=\columnwidth]{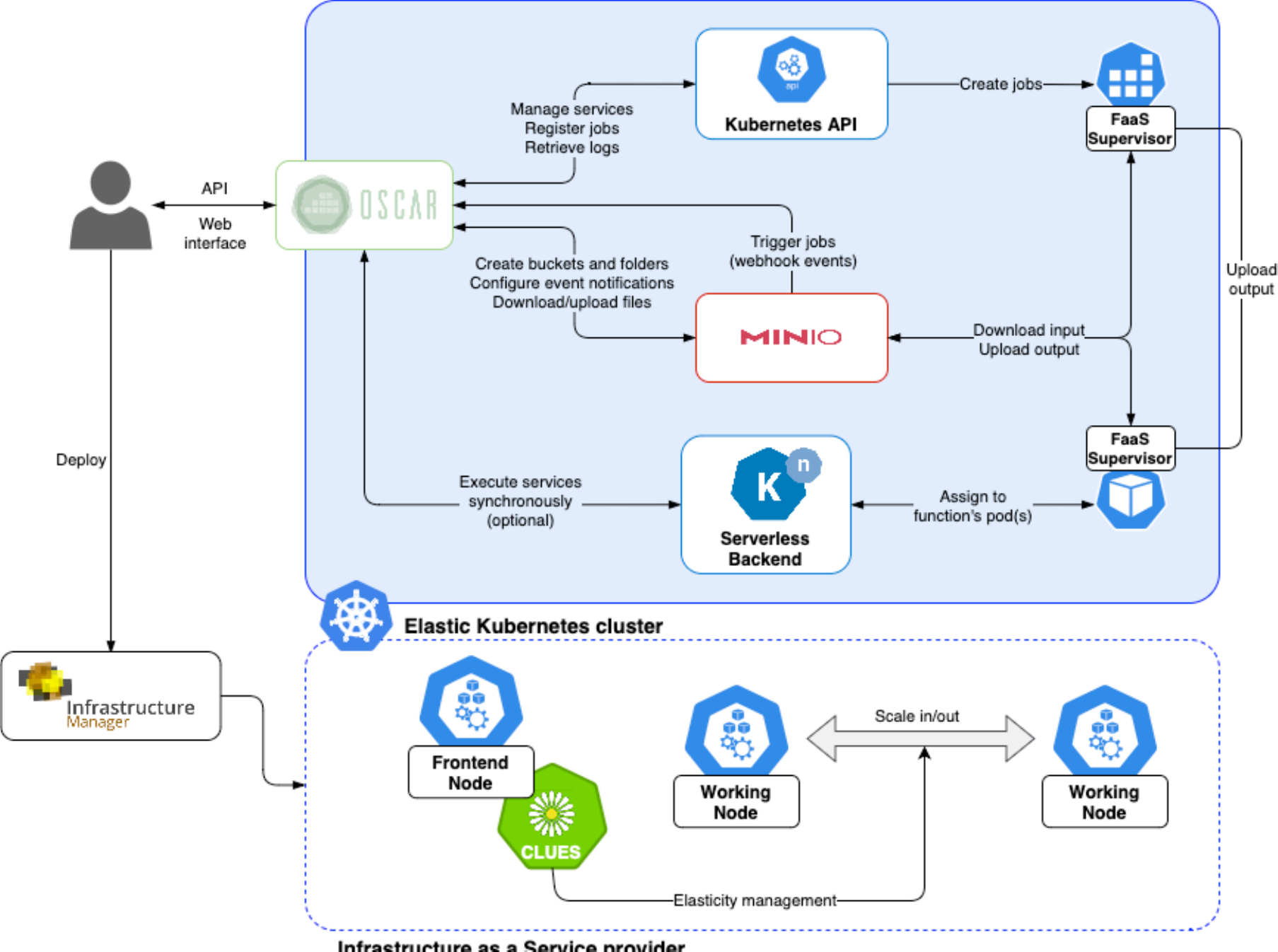}
  \caption{OSCAR and FaaS components including KNative integration.}
  \label{fig:oscar}
\end{figure}

In this section we explore the deployment of FaaS, using the OSCAR platform on top of Kubernetes. Oscar is designed for docker-based computational tasks on Kubernetes clusters across multi-cloud environments, making it an ideal choice for adaptable and scalable FaaS deployment. A cluster with OSCAR comprises several components, including technologies like CLUES for elasticity management, MinIO for high-performance distributed object storage, Knative as serverless platform, OSCAR Manager for API and service management, and OSCAR UI for an end-user graphical interface, as depicted in Figure \ref{fig:oscar}.  In terms of storage, OSCAR supports various external storage providers, including MinIO servers, Amazon S3, data access in the EGI Federated Cloud, and any WebDAV storage provider, such as dCache or StoRM.

We have chosen OSCAR as it works with KNative as FaaS service. OSCAR is also designed for data processing, as it triggers parallel function invocations through data events for file processing. In addition, KNative provides broker functionalities that can be customised. This broker offer a discoverable endpoint for event input (system details) and use triggers for event delivery (execution of functions). The broker serves as an intelligent intermediary between the pool of compute node resources and the ability to execute functions on various nodes. Its main function is to dynamically optimise the allocation of these functions/tasks to specific nodes, making informed decisions based on the current state of the system and other variables. This system state is collected by the broker and takes into account for each node a) availability and usage of computing units (CPU, GPU, ARM and others), b) access and availability of data collections, and c) system metrics in terms of access and connectivity bandwidth between nodes. With all this information, the Broker can use optimisation algorithms to make more intelligent decisions on how to execute the computing tasks more efficiently.


The next subsection summarise the installation process for OSCAR with basic components and then we include how to containerise and publish a function within for the FaaS and KNative.

\subsection{Deploying OSCAR and publishing functions in FaaS with OSCAR }

The OSCAR installation process consists of the following steps: a) installation of a Kubernetes cluster, deployment of Ingres-Nginx, deployment of MInIO and NFS on all nodes, deployment of KNative (for the FaaS service) and finally deploying the OSCAR components. All steps and a more detailed description can be found in the project repository \cite{Rodriguez_MOEA_for_SRC_2024}.

OSCAR allows the creation of serverless file processing services based on container images. These services require a user-defined script with the commands responsible for processing. The platform automatically mounts a volume in the containers, accesses the file that invokes the service, executes the user-defined script and loads the contents of the output folder. The basic instructions to run a filter on an image are depicted within project repository indicated previously. 

\section{MOEA developement}


The SRCNet comprises a network of nodes distributed globally, interconnected to form a fully connected graph. However, variations in resource availability, including occasional downtimes, necessitate periodic reevaluation of node connectivity and functional execution capabilities. The challenge entails devising an execution plan, given the current SRCNet structure, to optimize both execution time and energy consumption under specific conditions. This optimisation depends on system specifications and individual node characteristics, such as inter-node distance, bandwidth, and node information. Initially, the system's behavior will be theoretically modeled, rationalizing the selection of certain components of the model function (e.g., objective function and its domain) and the constraints imposed. While the solution may not encompass the entire solution space, practical guidelines will facilitate broader scope attainment. In practice, addressing the problem's extension remains a critical consideration. The objective is not to achieve the optimal solution but to derive a local solution enabling relatively efficient system request execution


\subsection{Problem formulation and model}

%

It is necessary to establish an initial approach to what is sought in this problem, in order to ultimately create a correct software implementation based on a series of choices that best suit the problem, such as support algorithms. We aim to find the minimum of a vector function, \( F: \Omega \to \mathbb{R}^L \) with \( \Omega \subset \mathbb{R}^P \) and \( L, P \in \mathbb{N} \). Initially, one might consider the classic problem of vector function optimisation, where it would be necessary to calculate the Jacobian of \( F \), \( \text{Jac}(F) \), the matrix of partial derivatives of the component \( F_i \) with respect to the coordinate \( x_j \), for \( i = 1, \ldots, L \) and \( j = 1, \ldots, P \). We confront our first problem: the prerequisite for $F$ to be differentiable, which does not have to be true. Moreover, considering the constraint functions, Lagrange multipliers could be used to calculate solutions according to them, which is computationally costly.


We want to find the minimum of the components of a vectorial function, in this context, the use of MOEAs may be a promising option to offer a solution to the problem. Hence, we might find a suboptimal execution plan, more likely with greater complexity of the function \( F \). In addition, MOEAs offers a good balance between computational efficiency and the objective to be achieved, ensuring that the algorithm does not exceed its execution time.

An approach has been developed that simulates the behavior of a single-node system, yielding results consistent with the desired outcome. Later on, it was extended to a multi-node system, making a generalisation in the function to be minimized at the time of implementation.

The developed model \( F \) was first thought for a single node of the system, but it generalizes to \( M \) nodes and \( P \) decision variables (dimension of the domain of \( F \)). However, it is implemented to optimize execution time and energy consumption, which means \( L = 2 \), but it can be seen that, although it implies a considerable extension of the algorithm, theoretically there is no problem in extending the function to \( L \) variables. On the other hand, despite being implemented with scalability in mind, it has been executed for \(P\)=3, that is number of CPU cores, number of GPU cores and number of ARM cores (one could say that \(P\) is the number of variables that, given certain constraints, the algorithm can alter to find the minimum, and and that is why the load on the node in the past few days, for example, has not been considered as decision variables).

\subsubsection{System information index}

The SRCNet will comprise geographically dispersed nodes sharing data, services and events. This creates a dynamic system where the status of nodes and the aforementioned components evolves over time. Consequently, the presence of a system information index becomes crucial, serving as a resource for the MOEA engine. Every time that the system change (be it the first launch, a node failure, a loss of some data block in a node, etc.), it is intended that the characteristics known to the algorithm be updated. Since the system is dynamically updated in response to changes that occur, the MOEA could be forced to recalculate suboptimal execution plans, allowing for some dynamism in the system.

On the other hand, the parameterisation necessary for the correct functioning of the algorithm, represented in a JSON style file, indicating the way the algorithm expects the information for a certain node, which we identify with \( id = Q \), with \( Q = 1, \ldots, M \), are the following:

\begin{itemize}
    \item \textit{ID}: Positive $int$ that identifies the node. In this case, $ID=Q$.
    \item \textit{NAME}: Name of the node, which could be the country it belongs to.
    \item \textit{CPU}: Positive $int$ indicating the number of available CPU cores in the node, regardless of the use it is being given.
    \item \textit{GPU}: Positive $int$ indicating the number of available GPU cores in the node, regardless of the use it is being given.
    \item \textit{ARM} or other architectures: Positive int indicating the number of available \textit{ARM} cores in the node, regardless of the use it is being given.
    \item \textit{LOAD}: Load of the node in the last $x$ days. It is a real number between $0$ and 1, with $0$ meaning it has been free, and $1$ meaning it has always been occupied with other functions.
    \item \textit{CONNECTIONS}: List of connections of node $Q$ with other nodes. If the data block is not in the node, it will be searched among its connections (direct or indirect) where to execute it. This process is detailed later.
    \item \textit{DATA}: List of data blocks/data collections available in each node for the execution of functions.
\end{itemize}

Additionally, for each node, there are a set of hardware features, and it is crucial to have information about the node's usage in the last $x$ days to provide a reliable approximation. Since the data will be distributed among the nodes, it will also be necessary to use a data structure to inform the MOEA of the location of the required data for the execution of a function. All of this is stored on the system information index.

\subsubsection{Solution representation}

As previously specified, the objective function \( F \) has its domain in a \(P\)-dimensional space and its range in an $L$-dimensional space. Therefore, the optimal solution will be the \(P\)-tuple \( (x_1, \ldots, x_P) \in \Omega \) such that \( F(x_1, \ldots, x_P) \leq F(y_1, \ldots, y_P) \) for all \( (y_1, \ldots, y_P) \in \Omega \).

\begin{figure}[htbp]
\centering
\includegraphics[width=0.5\textwidth]{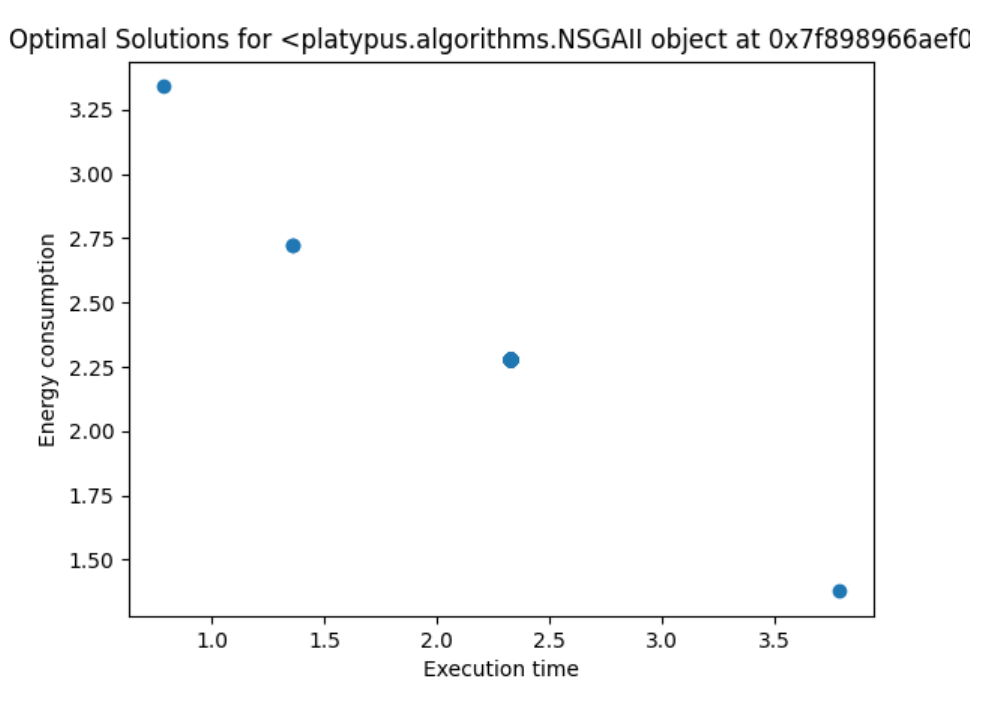}
\caption{One node behaviour.}
\label{fig:graphic1}
\end{figure}

If generalized to \( L \in \mathbb{N} \), the minimum of the function from a \(P\)-dimensional space is found in an \(L\)-dimensional space, as \( F \) is an \(L\)-dimensional function. To observe the cloud of points of possible solutions, we can only do so up to \( L\)=3, as computationally more dimensions cannot be visualized. Consequently, and since we want to verify that the behavior of the solutions is as expected (a direct relationship as it is a simple approximation), only the solution point cloud for \( L=2 \) and a node is shown in Fig~\ref{fig:graphic1}. As an example of the above, the following solution point cloud represents the behaviour for a node, in a much simpler but very restrictive approximation (since it is assumed that if GPU is used, the rest are not used, and the same applies to the other components). As will be seen later, the restrictions are relaxed, so a graph with a greater number of points is expected.

On the other hand, a solution (P-tuple associated with a point in the previous point cloud) is comprised of the following components (for a single-node system):
\begin{itemize}
    \item CPU cores: The number of CPU cores utilized within the node during an execution.
    \item GPU cores: The number of GPU cores utilized within the node during an execution.
    \item ARM cores: The number of ARM cores utilized within the node during an execution.
\end{itemize}

\begin{figure}[b]
\centering
\includegraphics[width=0.5\textwidth]{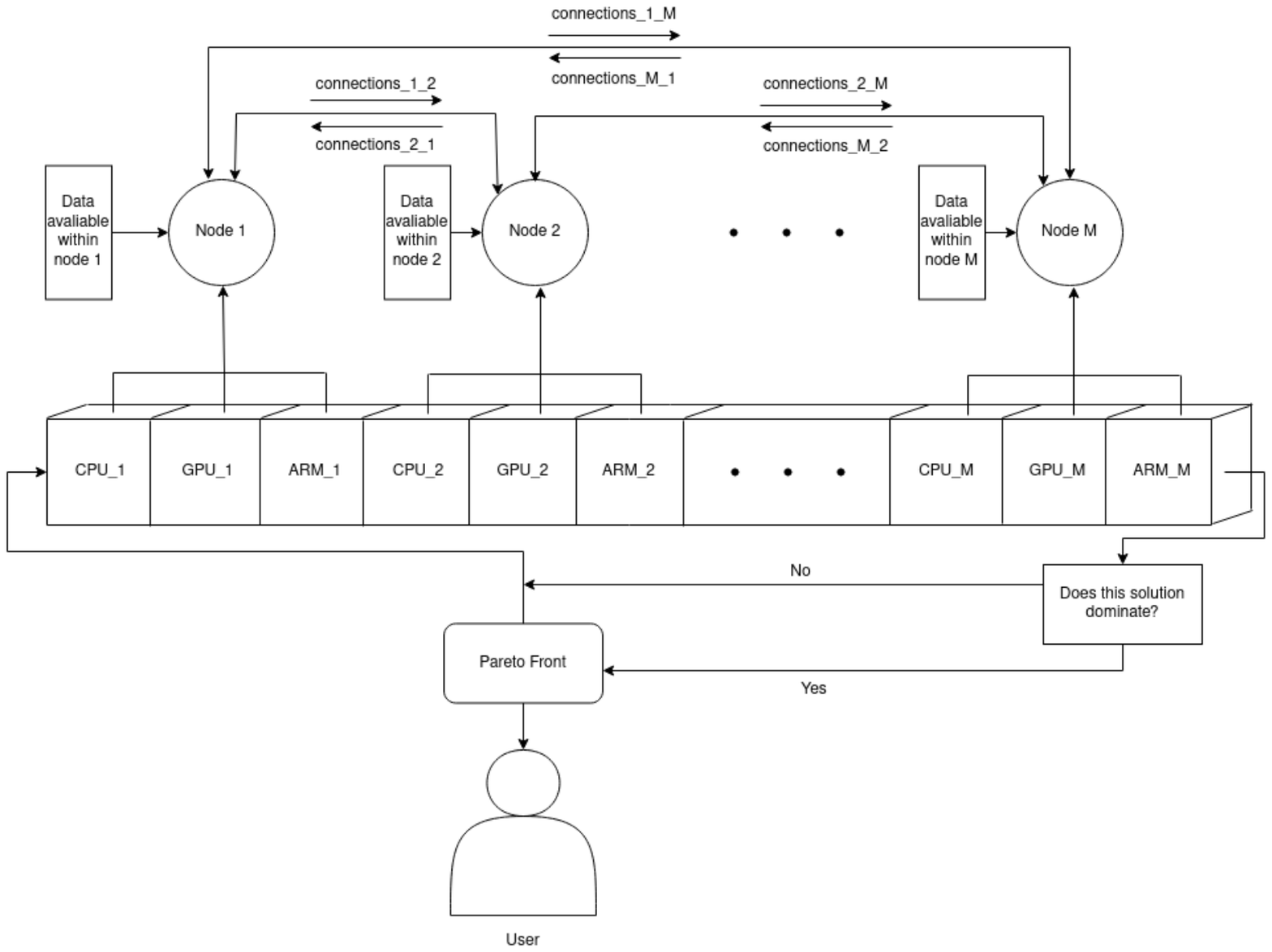}
\caption{Diagram depicting how the algorithm works and what components a solution has in order to obtain an execution plan.}
\label{fig:sol_repr}
\end{figure}

It is important to emphasize that its position in the solution vector has been demonstrated. The position of one of the decision variables corresponds to a certain characteristic of a node identified by its id. Later on, in data access, the importance of the order will be seen. 
Generally, in a system with \(M\) nodes, the situation is the same but multiplied by \(M\). However, the position is important because it represents the id of the node. In Fig~\ref{fig:sol_repr} is shown how a solution is represented.

\subsubsection{Objective functions and constraints}

Until now, the constraints only prevent a solution from being the origin of coordinates. That is, nothing from the system is used, which cannot occur. Although theoretically this is impossible, as the probability of this happening is $0$ in the dense space \( \mathbb{R} \), and even though it is highly improbable that this occurs, especially as the value of \( P \) increases generally making the probability tend to $0$, it is included because, however small the probability, it can occur due to the finite representation capacity of computers.

For \( P \) decision variables and \( M \) nodes, the algorithm implemented is:

\begin{algorithmic}
\State $vars \gets \text{List of Numbers}$

\Procedure{notValidIfOnlyZero}{}
    \For{$i \gets 0$ \textbf{to} $\text{len}(vars) - 1$ \textbf{step} $P$}
        \State $aux \gets \text{sumarPVariables}(vars)$
        \If{$aux = 0$}
            \State \textbf{return} $1.0$
        \EndIf
    \EndFor
    \State \textbf{return} $-1.0$
\EndProcedure
\end{algorithmic}

If required, additional constraint functions could be incorporated into the problem using a similar implementation as described above. For the implementation of the MOEA approach, we opted for the Platypus framework due to its high flexibility in handling such scenarios. However, it is worth noting that if the system is currently active, an restart would be necessary for the changes to take effect.These constraints could be denoted as \(g_i: \Omega \to \Lambda\), for all \(i=1,\ldots,B\), where \(B \in \mathbb{N}\) represents the number of constraints and \(\Lambda \subset \mathbb{R}\). We say a solution \(x \in \Omega\) is valid if $$g_i(x) \leq 0$$ for all \(i \in \{1,\ldots,B\}\) (note that \(\Omega\) is the solution space, the domain of \(F\)). In the proposed implementation, \(\Lambda = \{-1,1\}\) because the constraints added imply whether \(x\) is valid or not; there are no other factors, such as the degree to which a solution is valid, for example. Therefore, \(x\) is valid if \(g_i(x)=-1\) and it is not valid if \(g_i(x)=1\) for all \(i=1,\ldots,B\).

Regarding the objective function, the goal is to minimize both execution time and energy consumed by the system, so \( L=2 \). Thus, we have \( F:\Omega \to \mathbb{R}^2 \) defined as \( F(x_1,\ldots,x_P)=(F_1(x_1,\ldots,x_P),F_2(x_1,\ldots,x_P)) \), with this \( F_1 \) gives the execution time and \( F_2 \) the energy consumed.

Before continuing, a series of clarifications need to be made about how the objective function operates.

a) \textit{Randomness}. As in this first approximation, there are no execution data on the network, it is assumed that the data follow a normal distribution. They are not included as a decision variable because they are external to MOEA, just like the node load.

b) \textit{Access}. To access a decision variable, the following expression is proposed. If the system has \( M \) nodes, there are \( P \) decision variables per node, and I want to access the K-th decision variable of a node with id N, the expression for access if we call the array containing them \( vars \), would be
    \[
    varDecision = vars[P \cdot (N - 1) + (K - 1)]
    \]
Both the position of the decision variable and that \( vars \) is ordered are important. That is, if you have the array of P decision variables of node \( N \) ordered in a certain way, that order must be respected in the rest of the nodes, appending to the right the array of node \( N+1 \) and to the left that of node \( N-1 \).

c) \textit{Minimum Path}. Given the matrix of connections between nodes, if node ii needs a data block it doesn't have, it sends the rest of the function execution to the node that does have it. First, the list of nodes that have such a data block is obtained, and the one that involves a minimum path is determined using the Dykstra algorithm for the minimum path.

d) \textit{Parallelisation behavior according to the number of cores and technology used}. As an approach, the sequence of functions \( \left\{ \frac{1}{n^p} \right\} \) is proposed, where \( n \) is the number of cores and \( p \) the type of technology used. In the case of CPU and GPU, for example, more cores imply faster speed, with the GPU being faster than the CPU, so \( p_{\text{CPU}} < p_{\text{GPU}} \). This choice will be justified later for the case of execution time. As the behavior of energy consumed is inverse to that of time, the sequence \( \left\{ n^p \right\} \) is proposed, with a similar justification.

Let's suppose that \( p>0 \) is fixed. If \( n_1, n_2 \) are two amounts of cores such that \( n_1 < n_2 \), it is observed that if \( T \) is the execution time of a core, it holds that \[ \frac{1}{n_1^p} \cdot T > \frac{1}{n_2^p} \cdot T \] as expected (the greater the number of cores, the less the expected time assuming uniform or almost uniform distribution of work). This can be seen because asymptotically the sequence \( \left\{ \frac{1}{n^p} \cdot T \right\} \) tends to 0, that is, \[ \lim_{n \to \infty} \frac{1}{n^p} \cdot T = 0 \] Furthermore, the justification that the smaller the number of cores, the longer the execution time follows from \[ \lim_{n \to 0}  \frac{1}{n^p} \cdot T  = \infty \]

Now suppose that \( n \) is fixed and consider two technologies, 1 and 2, with 1 being slower than 2, let's say \( p_1 < p_2 \). If \( T \) is the base time, it holds that \( \frac{1}{n^{p_1}} \cdot T > \frac{1}{n^{p_2}} \cdot T \). Moreover, it is important to note that although they asymptotically tend to 0 if \( n \to \infty \), they do not converge at the same rate. If we solve \[ \frac{1}{n^{p_1}} = \frac{1}{n^{p_2}} \] we get \( n^{p_1} - n^{p_2} = 0 \), that is, we can make \[ n^{p_1} \cdot (1-n^{p_2 - p_1}) = 0 \] and since \( n \neq 0 \) then \( n^{p_2 - p_1} = 1 \). If \( p_2 - p_1 \in \mathbb{N} \), by the fundamental theorem of algebra we have \( p_2 - p_1 \) solutions. We are only interested in the solution \( n=1 \), as the rest are complex or negative solutions. From here it follows that if \( p_1 < p_2 \) asymptotically the sequence \( \left\{ \frac{1}{n^{p_1}} \right\} \) converges slower than the sequence \( \left\{ \frac{1}{n^{p_2}} \right\} \), so the approximation to distinguish the technology is accurate. It also follows that \[\frac{1}{n^{p_1}} < \frac{1}{n^{p_2}}  \text{ if }  n \in [0,1] \]

On the other hand, it is important to know that the chosen sequence for a certain \( n \) must be monotonically decreasing according to \( p \) for its entire domain. Certainly, this occurs with \( \left\{ \frac{1}{n^p} \right\} \) where
    \[
        \frac{d}{dp} \left[ \frac{1}{n^p} \right] = -p \cdot n^{-p-1} < 0 \quad \forall p>0, \forall n>0
    \]

\begin{figure}[ht]
\centering
\includegraphics[width=0.5\textwidth]{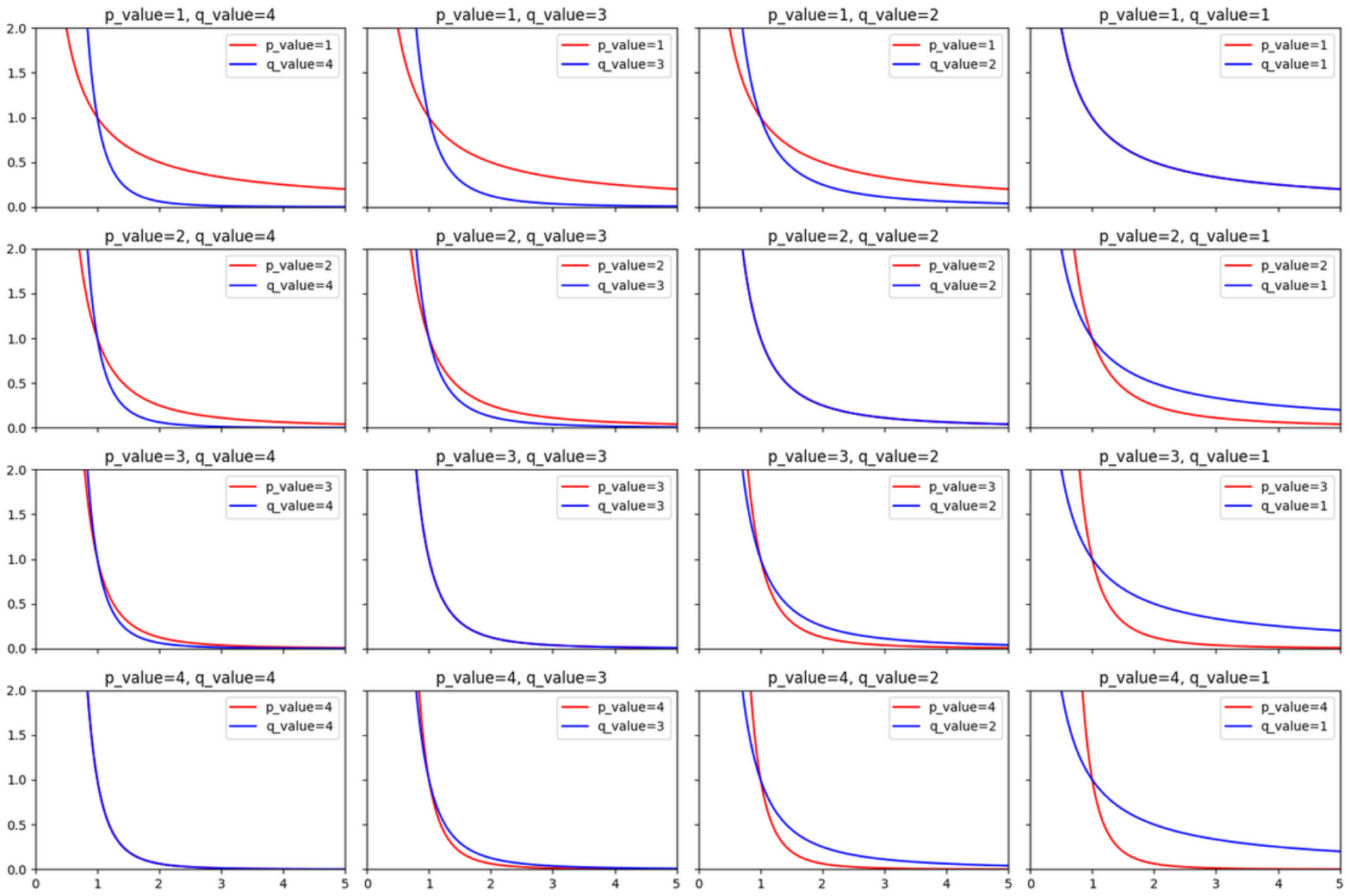}
\caption{Justification about $p$. One can observe that with a higher value of $p$, the graph converges more quickly for the cases of interest (number of cores greater than or equal to $1$).}
\label{fig:mi_etiqueta}
\end{figure}

Next, in Figure \ref{fig:mi_etiqueta} a graphical representation is shown according to the value of \( p \) for the considerations given above.

Returning to the objective function, we have two components \(F(x_1,...,x_P) = (F_1(x_1,...,x_P), F_2(x_1,...,x_P))\), where \(F_1\) is the execution time and \(F_2\) is the energy consumed, which are considered independent. However, a more realistic approximation would be \(F_2(x_1,...,x_P) = (G \circ F_1)(x_1,...,x_P)\) for some scalar function \(G: \Omega \to \mathbb{R^+}\) with \(\Omega \subset \mathbb{R^+}\), or \(F_2(x_1,...,x_P) = G(x_1,...,x_P, F_1(x_1,...,x_P))\) with \(G\) defined as \(G: \Omega \to \mathbb{R^+}\) where \(\Omega \subset \mathbb{R}^{P+\text{1}}\). The function is detailed according to the component we wish to model, assuming \(A\) is the set of data necessary for executing a function, with \(\#A\) being the number of necessary data points, \(connectionsMatrix\) the matrix of connections between nodes, with \(connectionsMatrix_{i,j} = -1.0\) if node \(i\) is not connected to \(j\), \(connectionsMatrix_{i,j} > 0\) if they are connected and \(i \neq j\), and \(connectionsMatrix_{i,j} = 0\) if \(i = j\). In addition, although initially all nodes are connected to each other, a solution is implemented for a non-connected network (as there may be cases where some connections fail), for this context we define the execution time and energy consumption. 

a) \textit{Execution Time}. Following the previously mentioned notation, it is represented by \(F_1\), which needs to be positive (as negative time cannot occur, and zero time is impossible). Given the node \textit{ID}, the execution plan, and the necessary data to execute a certain function, for each required data block, the shortest path to the node containing it is sought, returning $0$ if it is on the same node, a positive real value and the \textit{ID} to the node that has the data block accessed via a shortest path, or otherwise it is assumed to be not connected to the node holding those data (or they are not on the network at all) and is penalized, thus preventing the execution plan from being valid. This is based on \textit{Dykstra's} algorithm. Furthermore, as the goal is for it to continue executing on external nodes (bringing computation to the data), it is necessary to take into account the execution time on the foreign node. It is assumed that if it executes on another node, there are no dependencies with other nodes. That is, if node \(i\) sends to execute part of the function on node \(j\), only the time for the part of the function requiring the data block in \(j\) is extracted from this node. If more data blocks not in \(j\), say in a node \(k\), are needed, the call is repeated with said node. It is important to note that execution on the proposed node \(i\) is measured separately (that is, to ensure the node's load is only applied once per node).

On the other hand, assuming a uniform load distribution in the node, if \(hardware\) is the array of decision variables with the respective hardware components of a node, we define \(contribution_i\) as the contribution of each component to the workload, with value 

\[
contribution_i = \frac{hardware_i}{\sum_{j=0}^{j=\text{len}(\text{hardware})-1} hardware_j} 
\]

If \(T\) is the standard execution time, the expected time according to the \(hardware\) components will be 

\[
\sum_{j=0}^{j=\text{len}(\text{hardware})-1} \frac{1}{contribution_j^{p_j}} \cdot T 
\] 

ensuring that 

\[
\sum_{j=0}^{j=\text{len}(\text{hardware})-1} contribution_j = 1
\]    
Finally, to take into account the node's load over the last \(X\) days, the expected time for the execution of a function on node \(i\) will be \(time_i = (1+load_i) \cdot time\), where \(time\) is the expected time if \(load_i = 0\). In summary, with node i load noted as \(l_i\), and time of processing function noted as \(t_{\text{function}}\), the general expression would be, if node id is \(i\) and B is data within node \(i\):

    \begin{multline}
        F_{1,i}(x_1,...,x_P) = (1+l_i) \cdot \\ ( \sum_{j \in A-B}{t_{i,\text{minNode}_j} + 
         t_{\text{minNode}_j,i}} + \\ 
        F_{1,\text{minNode}_j}(x_1,...,x_P) + t_{\text{function}} )
    \end{multline}
    


b) \textit{Energy consumption}. Corresponding to \(F_2\) in the previous notation, a significant portion of it has a computation method analogous to that of execution time, relying on \textit{Dykstra} to estimate the minimum execution time on other nodes where the data block not in the same node is located, and from there the consumed energy is inferred taking into account the available hardware components. In this case, although the distribution is also uniform, the calculation method is different. That is, if \(hardware\) are the hardware components, part of the decision variables of the problem, associated with a node \(i\), \(contribution_j\) is obtained for \(j=0,...,\text{len}(\text{hardware})-1\) and the estimated energy consumption in the node is calculated as:

\[
\sum_{j=0}^{j=\text{len}(\text{hardware})-1} {contribution_j^{p_j}} \cdot T
\]

The general expression is analogous to that of time, but with the changes expressed above, where if for a certain node \(i\), we denote $B$ as the data within the node, and \(E_{\text{function}}\) denotes energy consumption in processing function, it would be:

\begin{multline}
            F_{2,i}(x_1,...,x_P) = (1+l_i) \cdot \\ ( \sum_{j \in A-B}{E_{i,\text{minNode}_j} + E_{\text{minNode}_j,i}} + \\F_{2,\text{minNode}_j}(x_1,...,x_P) +
            E_{\text{function}} )
\end{multline}

\subsection{Implementation}


The implementation was carried out using Python, leveraging the Platypus framework for Multi-Objective Evolutionary Algorithms (MOEA). Specifically, we employed the NSGA-II algorithm, including a population size of 100 and initial population generation using a Random Generator. For solution selection, we utilized a binary tournament selection approach, while the Simulated Binary Crossover (SBX) served as the recombination operator. Initial testing was conducted on a virtual machine (VM) provided within the computational infrastructure of the espSRC \cite{garrido2021toward}, which also included the setup of the Kubernetes platform necessary to support FaaS capabilities.

\section{Conclusion}

In this work, we want to emphasize the importance of designing a proposal that effectively addresses the challenge of bringing computation directly to the location of data, thus overcoming difficulties associated with moving large volumes of data. The adoption of the FaaS model emerges as a significant option, enabling the remote execution of functions in a distributed infrastructure. We have presented a preliminary proposal for optimizing execution plans for pipelines on FaaS within the context and architecture of the SRCNet using an approach based on MOEA. Our approach aims to facilitate decision-making in the DataMesh, optimizing computing operations on data with respect to critical aspects such as execution time and energy consumption. This methodology could be integrated within a Broker, enabling more efficient and adaptable management of distributed computing resources across the SRCNet architecture and marking a step towards efficiency and optimisation in large-scale astronomical data processing.

As future work, we propose an extension of the designed algorithm to include real metrics and node load data based on current loads, rather than relying on simulations or theoretical models as done so far. This will allow for more precise and adaptive decision-making, further enhancing the efficiency and responsiveness of the system in dynamic operational environments. Another important area for future research would be the detailed study of energy consumption of functions and operations on data, which would allow for more effective feeding of the MOEA algorithm and improving its performance in resource optimisation. Finally, a comprehensive evaluation of the integration and effectiveness of the Broker using MOEA is required to measure the efficacy and practical impact of the approach proposed in this work.


\section*{Acknowledgment}

We acknowledge financial support from the grant CEX2021-001131-S funded by MCIU/AEI/ 10.13039/501100011033 and from the grant DOC01497 funded by the  Consejería de Transformación Económica, Industria, Conocimiento y Universidades de la Junta de Andalucía and by the Operational Program ESF Andalucía 2014-2020. We acknowledge financial support from the grant TED2021-130231B-I00 funded by MCIU/AEI/ 10.13039/501100011033 and by the “European Union NextGenerationEU/PRTR”. We also acknowledge financial support from the grant  PID2021-123930OB-C21 funded by MCIU/AEI/ 10.13039/501100011033 and by “ERDF A way of making Europe”.



%

\bibliographystyle{IEEEtran}
\bibliography{bibliog}

\end{document}